# Analysis of Local Methane Emissions Using Near-Simultaneous Multi-Satellite Observations: Insights from Landfills and Oil-Gas Facilities

Alvise Ferrari
*School of Aerospace Engineering*
Sapienza University
Rome, Italy
*GMATICS S.r.l.,* Rome, Italy
alvise.ferrari@uniroma1.it

Giovanni Laneve
*School of Aerospace Engineering*
Sapienza University
Rome, Italy
giovanni.laneve@uniroma1.it

Raul Alejandro Carvajal Tellez
*School of Aerospace Engineering*
Sapienza University
Rome, Italy
raulalejandro.carvajaltellez@uniroma1.it

Valerio Pampanoni
*School of Aerospace Engineering*
Sapienza University
Rome, Italy
valerio.pampanoni@uniroma1.it

Simone Saquella
*DIAEE*
Sapienza University,
Rome, Italy
simone.saquella@uniroma1.it

Rocchina Guarini
*Italian Space Agency (ASI)*
Rome, Italy
rocchina.guarini@asi.it

*Abstract*— Methane ($CH_4$) is a potent greenhouse gas, and its detection and quantification are crucial for mitigating the greenhouse effect. This study presents a comparative analysis of methane emissions observed using near-simultaneous observations from hyperspectral imaging spectrometers hosted aboard different satellite platforms (PRISMA, EnMAP, EMIT and GHGSat). Methane emissions from oil and gas facilities and landfills are analyzed to evaluate the consistency and precision of the sensors and temporal variability of the source. Landfills, characterized by diffuse and stable emissions, and dynamic oil and gas facilities, subject to operational variability, provide contrasting use cases for emission monitoring. Emission rates are quantified using the Integrated Mass Enhancement (IME) model and validated across satellites with overlapping acquisitions. This study highlights the advantages and limitations of each satellite system, emphasizing the critical role of multi-sensor integration in bridging temporal and spatial observation gaps. Insights derived here aim to enhance global methane monitoring frameworks and guide future satellite design for improved emission quantification.

## I. Introduction

In recent years, there has been a significant increase in interest in the detection and measurement of local methane ($CH_4$) emissions, driven primarily by the availability of various satellites capable not only of identifying the presence of methane but also of quantifying its concentration in plumes. Methane detection methodologies, initially validated with airborne missions like AVIRIS and AVIRIS-NG [1,2,3], have since been successfully adapted for satellite instruments such as PRISMA, EMIT, and EnMAP [5,6,7,8]. These missions have demonstrated their effectiveness in local methane emissions remote sensing, offering high spatial resolution (30 m to 60 m). In the commercial domain, the first satellite mission dedicated to $CH_4$ detection, GHGsat, pioneered methane-specific monitoring and measurement [9].

Currently, the dominant methodology for $CH_4$ retrieval is based on a formulation of the clutter-matched filter (CMF) [12], which uses a unit target spectrum derived from regression analysis between at-sensor radiance spectra simulated by varying methane concentrations, with other geometric and atmospheric parameters tailored to the specific scene, enabling both detection and quantification. The effectiveness of PRISMA in methane detection has been demonstrated [5, 6], while EnMAP has recently been studied for $CH_4$ retrieval, showing advantages compared to PRISMA [8]. EMIT's ability to detect methane emissions with high accuracy, leveraging its advanced spectral calibration and high Signal-to-Noise Ratio (SNR) values in the ShortWave Infrared (SWIR) region has been recently highlighted by Thompson et al. (2024) [10].

Recently, the integrated use of multi-source satellite observations has enabled more continuous monitoring of globally significant methane-emitting sites. For instance, Zhang et al. (2024) [11] combined TROPOMI's broad detection capabilities with the high-resolution observations of EMIT, EnMAP, and PRISMA to quantify emissions from solid waste landfills.

Despite the advancements in methane monitoring using multi-satellite observations, a key gap remains in the lack of comparative analyses of these instruments over the same points of interest (POIs) with acquisitions taken in very close temporal proximity. This study aims to address this gap by targeting methane emissions from both landfills and oil and gas facilities. While oil and gas facilities are often dynamic emitters influenced by operational variations, landfills are relatively more stable emitters within short timeframes and under stable atmospheric conditions. By comparing measurements from satellites with near-simultaneous overpasses, this study aims to evaluate the consistency and precision of methane emission quantifications, offering insights into temporal variability and potential methodological harmonization for future monitoring efforts.

## II. METHODS

### A. Clutter Matched Filter and Target Spectrum

Methane exhibits strong absorption of electromagnetic radiation in specific SWIR regions, centered around 1650 nm and 2300 nm. These absorption features enable the detection and quantification of methane concentrations by analyzing solar radiance reflected from the Earth's surface and captured by hyperspectral sensors aboard satellites or aircraft. In this study, we adopt a matched filter (MF) based methodology, which models the at-sensor radiance measured by the hyperspectral sensor as a linear combination of three components: the signal associated with the methane enhancement, scaled by its strength (MF score), an average background, and a noise or clutter term that accounts for both sensor noise and scene background variability. The background is modeled as a multivariate Gaussian distribution characterized by a mean $\mu$ and covariance matrix $\Sigma$. Deviations from this background are associated with methane enhancements through a target spectrum $t$, representing the absorption spectrum equivalent to a unit methane concentration. In each pixel of an image, methane concentration enhancements $\Delta X_{CH4}$ are directly estimated using the formulation:

$$\Delta X_{CH4} = \frac{(x-\mu)^T \Sigma^{-1} t}{(t^T \Sigma^{-1} t)} \quad (1)$$

where $x$ is the observed radiance spectrum at each pixel. this direct estimation avoids intermediate MF score calculations and simplifies the retrieval process. The target spectrum $t$ [ppm·m$^{-1}$] is derived using MODTRAN6 radiative transfer simulations with all parameters tailored to the specific acquisition, as detailed in [6]. In this process, scene-specific at-sensor radiances are computed for discrete methane enhancement values ranging from 0 to 64,000 ppm·mm. For each simulated wavelength, $t$ is derived by performing a linear regression between the natural logarithm of the radiance and the methane enhancement values. To ensure compatibility with the satellite sensor, the resulting target spectrum is then convolved with the sensor's spectral response function. This scene-specific target generation approach enhances retrieval sensitivity, and robustness and to automate and speed-up target spectrum generation, a Lookup Table (LUT) replaces real-time MODTRAN6® simulations with precomputed at-sensor radiances that account for variations in atmospheric and geometric parameters such as sensor altitude, water vapor content, ground elevation, solar zenith angle, and methane concentration, as described by [4,6].

### B. Estimation of Emission Rate

Methane (CH$_4$) concentrations in remote sensing are often expressed in ppm·m (parts per million multiplied by meters) to represent integrated column measurements. For applications requiring volumetric concentrations (ppb), a conversion factor of 0.125 is used, based on an assumed 8 km atmospheric scale height and uniform vertical distribution of methane [5].

*1) Plume segmentation:*

Accurate IME calculation requires isolating the plume from the background and ensuring only relevant pixels are included. The emission flow rate (Q) is derived by manually delineating a polygon around the plume; ultra-high-resolution imagery (e.g., WorldView-3) is used to identify and exclude false-positive sources or artifacts. Emissions from heterogeneous surfaces are discarded if artifacts hinder clear plume identification.

*2) IME calculations:*

The calculation of the emission flow rate (Q), based on the pixel-by-pixel methane column enhancements ($\Delta$XCH4) in the plume, is performed using the Integrated Mass Enhancement (IME) model [18,19]. The IME, expressed in kilograms, represents the total excess mass of methane within the plume and is computed as:

$$IME = k \cdot \sum_{i=1}^{n_p} \Delta X_{CH4}(i) \quad (2)$$

where $n_p$ is the number of pixels in the plume, and k is a scaling factor scaling used to convert the methane enhancement values ($\Delta X_{CH4}$, in ppb) into kilograms. The scaling factor k is calculated as:

$$k = \frac{M_{CH_4}}{M_a} \cdot \Sigma_a \cdot A_p \cdot 10^{-9} \quad (3)$$

where $M_{CH_4}$ (g/mol) is the molar mass of methane, $M_a$ (g/mol) the molar mass of dry air, $\Sigma_a$ (kg/m$^2$) is the column-integrated dry air mass, $A_p$ (m$^2$) is the area of a pixel and $10^{-9}$ converts ppb to fractional concentration. This calculation assumes Avogadro's law and uses standard atmospheric conditions.

*3) Flow rate computation:*

The flow rate Q [kg/h] is then calculated as:

$$Q = \frac{IME \cdot U_{eff}}{L} \quad (4)$$

where $U_{eff}$ is the effective wind speed, and L represents the plume length scale, defined as the square root of the plume-mask area. The effective wind speed $U_{eff}$ is estimated from the 10 m height wind speed $U_{10}$ (m/s), using large-eddy simulations specifically designed for the spatial resolution and retrieval accuracy of $\Delta X_{CH4}$ consistent with each satellite's characteristics, and modelling the emitter as area sources, as reported in [11], resulting in a linear dependency of $U_{eff}$ from $U_{10}$:

$$PRISMA: U_{eff} = 0.37 \cdot U_{10} + 0.70$$

$$EnMAP: U_{eff} = 0.37 \cdot U_{10} + 0.69$$

$$EMIT: U_{eff} = 0.45 \cdot U_{10} + 0.67$$

The accuracy of the measurements and the propagation of uncertainties in Q, including contributions from wind speed and IME, will be addressed in greater detail in a future, more comprehensive study.

## III. MATERIALS

In this section, we provide a concise overview of the fundamental aspects of the input data and the algorithmic implementations specific to methane concentration product generation for each sensor.

*A. PRISMA and EnMAP:*

PRISMA, developed by the Italian Space Agency (ASI), and EnMAP, managed by the German Aerospace Center (DLR), are hyperspectral imaging satellites with similar instrument designs. Both missions cover the same spectral range, and share a ground sampling distance (GSD) of 30 m. We utilized L1B and L1 radiometrically calibrated radiance data for EnMAP and PRISMA, respectively. The algorithms used for processing these data, based on the CMF, are available as containerized software in the online repositories [PRISMA-CH4](#) [13] and [EnMAP-CH4](#) [14], with implementation details for PRISMA also described in [6]. In this specific implementation, matched filters for PRISMA are computed on a column-wise basis to correct for the spectral smile effect, while the current implementation for EnMAP does not yet include smile correction. A version that accounts for smile-effect will be released in future repository updates of the EnMAP processing code. EnMAP and PRISMA satellites are similar hyperspectral imaging systems, but they exhibit critical differences in spectral and radiometric performance that affect their effectiveness in $CH_4$ detection. Based on findings by Roger et al. (2024) [8], EnMAP's narrower full-width at half-maximum (FWHM) of ~7.8 nm in the 2300-nm methane absorption window, compared to PRISMA's broader 10 nm, enables it to resolve methane absorption features more effectively, reducing the influence of background variability. This advantage outweighs PRISMA's higher spectral sampling density (SSD), which provides more spectral bands within the 2300-nm window (47 bands for PRISMA vs. 43 for EnMAP) but is less critical for methane detection. Additionally, EnMAP demonstrates a SNR approximately double that of PRISMA in the 2300nm window, resulting in reduced retrieval noise and enhanced detection capability, particularly in low-radiance conditions. EnMAP also shows lower sensitivity to spectral smile effects (1.3 nm across-track variation compared to PRISMA's 2.8 nm), ensuring more uniform spectral calibration across the image. However, EnMAP does exhibit more pronounced striping, necessitating column-wise corrections during processing, though this is less disruptive than PRISMA's greater vulnerability to spectral smile. These factors combine to give EnMAP a distinct advantage in methane retrieval, with lower retrieval noise (10.4 ppb for EnMAP vs. 21.6 ppb for PRISMA [8]) and superior separation of methane signals from background features.

*B. EMIT*

The EMIT instrument, developed as part of NASA's Earth Venture-Instrument Program, is a Dyson imaging spectrometer deployed on the International Space Station (ISS) to map soil mineral composition globally. With a spectral range spanning from 380 to 2500 nm and a bandwidth of approximately 7 nm, EMIT collects radiance data across a 75-km swath with a 60-m GSD. In addition to its primary objective of studying mineral dust, EMIT has proven effective for detecting greenhouse gases, such as methane ($CH_4$). The EMIT L2B Methane Enhancement Data product quantifies $CH_4$ enhancements at the pixel level using a CMF approach [15], employing a column-wise derived spectral covariance matrix to decorrelate background noise [16]. The resulting $CH_4$ enhancement is expressed in parts per million meter (ppm·m). False positives are minimized by excluding pixels influenced by clouds, water, or high radiance due to flares. The algorithm is highly sensitive to subtle methane signals, leveraging EMIT's high spectral resolution, high SNR, and wide swath for robust greenhouse gas monitoring. The methodology and algorithms are detailed further in the EMIT GHG repository [17].

*C. GHGSat*

GHGSat is a commercial constellation of 12 satellites (as of this paper's submission), equipped with custom Fabry–Perot imaging spectrometers targeting methane ($CH_4$) absorption lines near 1.65 μm, with a nominal GSD of ~50 m. The proprietary data processing algorithm, broadly described by Jervis et al. [9], employs a "differential" approach similar to DOAS (Differential Optical Absorption Spectroscopy) [1]. Top-of-atmosphere radiances are normalized against a "clean-air" background spectrum (e.g., upwind of emitters), with broad spectral trends (e.g., from surface reflectance or aerosol scattering) removed via polynomial fitting. A second-order polynomial captures slow signal variations, isolating the $CH_4$ absorption signature, which is compared to high-resolution absorption cross-sections from radiative transfer models such as MODTRAN. This produces column-enhancement maps (ppm·m). GHGSat integrates these maps with meteorological data to derive emission fluxes (e.g., kg/h), likely using the IME method [18,19], enabling high-resolution methane monitoring with low revisit times. We utilized GHGSat's L2 $CH_4$ concentration maps [ppb], ensuring compatibility with our analysis framework.

IV. RESULTS AND DISCUSSION

Several closely timed acquisitions were identified among GHGSat, PRISMA, EnMAP, and EMIT, enabling a comparative analysis of closely spaced temporal observations. Having access to all currently available hyperspectral sensors capable of measuring local methane emissions, we constructed a robust dataset for evaluating estimates from near-simultaneous acquisitions. Two high-interest cases are presented: the Buenos Aires landfill, one of the greatest methane-emitting landfills globally, and the area surrounding Kamishlidza, Turkmenistan, known for its gas compression stations. The Buenos Aires landfill represents a critical site for assessing emissions from diffuse and large-scale sources, while Kamishlidza, with its homogeneous background, provides an ideal environment for isolating and evaluating the inherent capabilities of the instruments in methane plume detection.

*A. Case 1: EnMAP – GHGsat – EMIT of 2024/01/12*

The Buenos Aires landfill case study (Fig. 1) offers an interesting opportunity for analysis due to the availability of three temporally close satellite acquisitions: GHGSat (2024/01/12 at 14:45:16 UTC), EnMAP (2024/01/12 at 14:46:53 UTC), and EMIT (2024/01/12 at 18:59:17 UTC). The near simultaneity of GHGSat and EnMAP, with a time difference of just 1 minute and 37 seconds, allows for a direct and particularly valuable comparison of the observed methane emissions. In this case, emission fluxes were calculated using the IME method, with a wind speed of 6.7 m/s at a direction of approximately 90° (source:

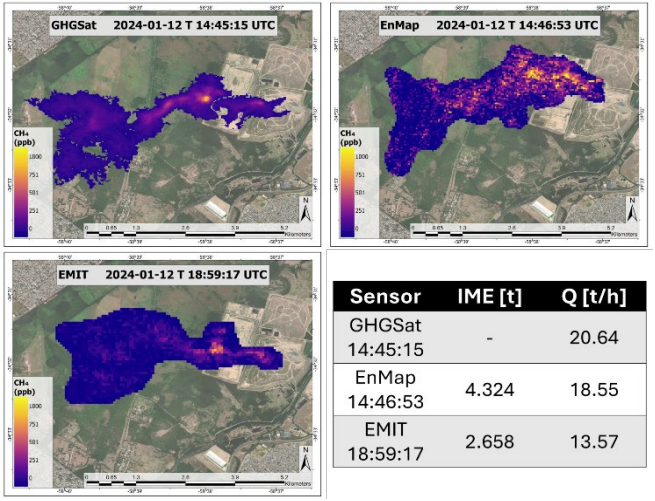

Figure 1- Buenos Aires landfill case study

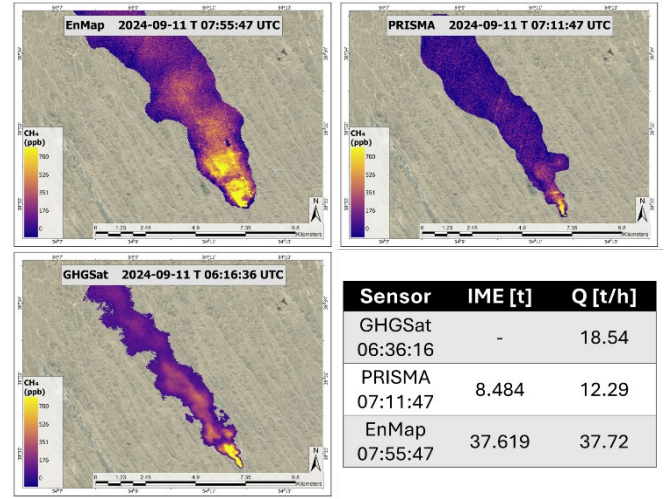

Figure 2 - Kamishlidza O&G facility case study

OpenWeather). GHGSat identified two distinct emission sources within the landfill area, producing two separate plumes that, given the wind direction and source configuration, tend to merge into a single plume. The combined flux estimated for these two emissions by GHGSat was 20.637 kg/h (14.958 kg/h and 5.679 kg/h, respectively), a significant value.

EnMAP's acquisition, performed just 1 minute and 37 seconds later (around 11:45 local time), shows excellent agreement: using our algorithm and manually defining the plume's extent in a manner similar to GHGSat, the flux calculation via IME yields a value of 18.55 t/h, very close to GHGSat's total estimate. This consistency between two different sensors observing the same area almost simultaneously suggests reliable flux estimates, although it should be noted that the images are not perfectly co-registered, and small discrepancies in pinpointing emission sources may exist. The situation is different in the case of EMIT, whose acquisition occurred over four hours later (around 16:00 local time), under different atmospheric and illumination conditions. In this case, the calculated flux was approximately 13.57 t/h. The lower flux is not necessarily due to instrumental or methodological limitations, nor can a real variation in emissions and environmental conditions over time be excluded. Despite the interpolation of the target spectrum in the LUT, the significant change in zenith angle may have influenced the measurement, making EMIT less able to capture the situation observed in the morning by EnMAP and GHGSat.

*B. Case 2: GHGSat - PRISMA – EnMAP of 2024/09/11*

This case study (Fig.2) examines methane emissions detected from a compression station in Kamishlidza, Turkmenistan, using three distinct satellite acquisitions: GHGSat at 06:36:16 UTC on 2024/09/11, PRISMA at 07:11:43 UTC, and EnMAP 45 minutes later at 07:55:47 UTC. Emission fluxes were calculated using the IME method with a wind speed of 2.7 m/s (source: OpenWeather). The three images show significant differences in plume intensity: GHGSat, the first to pass, detected an emission flux of 18.54 t/h. About 35 minutes later, PRISMA measured a flux of 12.29 t/h, and finally, EnMAP, 45 minutes after PRISMA, observed a much higher flux of 37.72 t/h, indicating a substantial emission. Despite similar zenith viewing angles (13.63° for PRISMA and 14.7° for EnMAP), the EnMAP output shows a more concentrated plume compared to PRISMA and GHGSat. This discrepancy can be attributed to temporal factors, such as variations in emissions within the 45-minute interval. From this case, it is not possible to draw definitive conclusions, as the flux may have first decreased and then suddenly increased. However, it should also be noted that EnMAP's signal-to-noise ratio (SNR) is about twice that of PRISMA, significantly reducing radiometric noise and improving estimation accuracy. While PRISMA provides denser spectral sampling (SSD), the higher SNR and lower sensitivity to spectral smile effects give EnMAP a significant advantage under the observed conditions. Therefore, PRISMA underestimation of the flux cannot be ruled out.

## V. Conclusions

This study demonstrates the potential of multi-satellite observations for accurate and consistent methane emission monitoring, utilizing all currently accessible hyperspectral sensors capable of $CH_4$ detection, including PRISMA, EnMAP, EMIT, and GHGSat. Despite inherent differences in sensor design and processing implementations, results reveal a high degree of consistency between instruments when observations are acquired within a short timeframe. However, discrepancies observed in some cases underscore the importance of understanding temporal variability in emissions, as well as the influence of sensor-specific factors such as SNR, view zenith angle, and atmospheric conditions. The findings support the integration of multi-sensor data to bridge spatial and temporal gaps and enhance global monitoring efforts. Future work will focus on better characterizing and reconciling inter-instrument differences to move toward a harmonized multi-satellite framework for methane monitoring, addressing algorithmic disparities and expanding the analysis to additional near-simultaneous acquisitions. By advancing our understanding of methane emissions through synergistic satellite observations, this study contributes to the development of robust frameworks for global greenhouse gas monitoring and mitigation.

ACKNOWLEDGMENTS

This research was supported by the Italian Space Agency ASI within the CLEAR-UP project Contract/Agreement n. N. 2022-16-U.0 CUP n. F83C22000780005.

We acknowledge the European Space Agency (ESA) Third-Party Mission (TPM) program for granting access to GHGSat acquisitions (Proposal ID: PP0100261).

REFERENCES

[1] Thorpe, A.K., Frankenberg, C. and Roberts, D.A., 2014. Retrieval techniques for airborne imaging of methane concentrations using high spatial and moderate spectral resolution: application to AVIRIS. *Atmospheric Measurement Techniques*, *7*(2), pp.491-506.

[2] Thorpe, A.K., Roberts, D.A., Bradley, E.S., Funk, C.C., Dennison, P.E. and Leifer, I., 2013. High resolution mapping of methane emissions from marine and terrestrial sources using a Cluster-Tuned Matched Filter technique and imaging spectrometry. Remote Sensing of Environment, 134, pp.305-318.

[3] Foote, M.D., Dennison, P.E., Thorpe, A.K., Thompson, D.R., Jongaramrungruang, S., Frankenberg, C. and Joshi, S.C., 2020. Fast and accurate retrieval of methane concentration from imaging spectrometer data using sparsity prior. IEEE Transactions on Geoscience and Remote Sensing, 58(9), pp.6480-6492.

[4] Foote, M.D., Dennison, P.E., Sullivan, P.R., O'Neill, K.B., Thorpe, A.K., Thompson, D.R., Cusworth, D.H., Duren, R. and Joshi, S.C., 2021. Impact of scene-specific enhancement spectra on matched filter greenhouse gas retrievals from imaging spectroscopy. Remote Sensing of Environment, 264, p.112574.

[5] Guanter, L., Irakulis-Loitxate, I., Gorroño, J., Sánchez-García, E., Cusworth, D.H., Varon, D.J., Cogliati, S. and Colombo, R., 2021. Mapping methane point emissions with the PRISMA spaceborne imaging spectrometer. Remote Sensing of Environment, 265, p.112671.

[6] Ferrari, A., Laneve, G., Pampanoni, V., Carvajal, A. and Rossi, F., 2024, July. Monitoring Methane Emissions from Landfills Using Prisma Imagery. In *IGARSS 2024-2024 IEEE International Geoscience and Remote Sensing Symposium* (pp. 3663-3667). IEEE.

[7] Nesme, N., Marion, R., Lezeaux, O., Doz, S., Camy-Peyret, C. and Foucher, P.Y., 2021. Joint use of in-scene background radiance estimation and optimal estimation methods for quantifying methane emissions using prisma hyperspectral satellite data: Application to the korpezhe industrial site. Remote Sensing, 13(24), p.4992.

[8] Roger, J., Irakulis-Loitxate, I., Valverde, A., Gorroño, J., Chabrillat, S., Brell, M. and Guanter, L., 2024. High-resolution methane mapping with the EnMAP satellite imaging spectroscopy mission. IEEE Transactions on Geoscience and Remote Sensing.

[9] Jervis, D., McKeever, J., Durak, B.O., Sloan, J.J., Gains, D., Varon, D.J., Ramier, A., Strupler, M. and Tarrant, E., 2021. The GHGSat-D imaging spectrometer. Atmospheric Measurement Techniques, 14(3), pp.2127-2140.

[10] Thompson, D.R., Green, R.O., Bradley, C., Brodrick, P.G., Mahowald, N., Dor, E.B., Bennett, M., Bernas, M., Carmon, N., Chadwick, K.D. and Clark, R.N., 2024. On-orbit calibration and performance of the EMIT imaging spectrometer. Remote Sensing of Environment, 303, p.113986.

[11] Zhang, X., Maasakkers, J.D., Roger, J., Guanter, L., Sharma, S., Lama, S., Tol, P., Varon, D.J., Cusworth, D.H., Howell, K. and Thorpe, A., 2024. Global identification of solid waste methane super emitters using hyperspectral satellites.

[12] Theiler, J., Foy, B.R. (2006). Effect of Signal Contamination in Matched-Filter Detection of the Signal on a Cluttered Background. IEEE Geosci. Remote Sens. Lett., 3, 98–102.

[13] A. Ferrari, "PRISMA-CH4", github repository, [Online], Available at: https://github.com/AlFe23/PRISMA-CH4

[14] A. Ferrari, "EnMAP-CH4", github repository, [Online], Available at: https://github.com/AlFe23/EnMAP-CH4

[15] Jet Propulsion Laboratory (2023). EMIT L2B Greenhouse Gas Data Product User Guide. Jet Propulsion Laboratory, California Institute of Technology. Available at https://lpdaac.usgs.gov/products/emitl2bch4enhv001/.

[16] Brodrick, P. G., Thorpe, A. K., Villanueva-Weeks, C. S., Elder, C., Fahlen, J., & Thompson, D. R. (2023). EMIT Greenhouse Gas Algorithms: Greenhouse Gas Point Source Mapping and Related Products - Theoretical Basis. Jet Propulsion Laboratory, California Institute of Technology. Available at https://lpdaac.usgs.gov/products/emitl2bch4enhv001/.

[17] Brodrick, P. G. et al., "emit-sds/emit-ghg: Mapping of greenhouse gases with EMIT. github repository, [Online], Available at: https://github.com/emit-sds/emit-ghg

[18] C. Frankenberg et al., "Airborne methane remote measurements reveal heavy-tail flux distribution in four corners region," Proc. Nat. Acad. Sci. India A, Phys. Sci., vol. 113, no. 35, pp. 9734–9739, Aug.

[19] D. J. Varon et al., "Quantifying methane point sources from fine-scale satellite observations of atmospheric methane plumes," Atmos. Meas. Techn., vol. 11, no. 10, pp. 5673–5686, Oct. 2018.

[20] D. H. Cusworth et al., "Potential of next-generation imaging spectrometers to detect and quantify methane point sources from space," Atmos. Meas. Techn., vol. 12, no. 10, pp. 5655–5668, Oct. 2019.